\documentclass[twocolumn,showpacs,preprintnumbers,amsmath,amssymb]{revtex4}
\usepackage{graphicx}
\usepackage{float}
\usepackage[usenames,dvipsnames]{color}
\begin{document}
   \title{Solving the Cusp-Core Problem with a Novel Scalar Field Dark Matter}
   \author{Kung-Yi Su$^{1,3}$}
     \email{b95202049@ntu.edu.tw}
   \author{Pisin Chen$^{1,2,3,4}$}
     \email{pisinchen@phys.ntu.edu.tw}
     \affiliation{%
1. Department of Physics, National Taiwan University, Taipei, Taiwan 10617\\
2. Graduate Institute of Astrophysics, National Taiwan University, Taipei, Taiwan 10617\\
3. Leung Center for Cosmology and Particle Astrophysics, National Taiwan University, Taipei, Taiwan 10617\\
4. Kavli Institute for Particle Astrophysics and Cosmology, SLAC National Accelerator Laboratory, Stanford University, Stanford, CA 94305, U.S.A.
}%

\date{\today}

\begin{abstract}
Matos, Guzm\'{a}n and Nu\~{n}ez (MGN) proposed a model of galactic halo based on an exponential-potential scalar field that could induce a rotation curve that is constant for all radii. We demonstrate that with suitable boundary conditions, such scalar field dark matter (SDM) can not only produce the observed constant rotation curve at large radii but also give rise to the correct power-law scaling near the galactic core region. This solves the existing cusp-core problem faced by the conventional cold dark matter (CDM) model.
\end{abstract}

\maketitle
\section{Introduction}

The nature of dark matter has always been an important issue in the study of cosmology. The Cold Dark Matter (CDM) model, which assumes the dark matter as weakly interacting massive particles (WIMPs), has been very successful in explaining the CMB anisotropies and the large scale structures in the Universe. However, it also has encountered several challenges such as the missing satellites problem \cite{0004-637X-522-1-82,Moore:2005jj,Diemand:2006ik} and the cusp-core problem \cite{Mikheeva:2007kp,Diemand:2005vz}.

The cusp-core problem arises from the mismatch between the density profile obtained from the simulations based on the CDM model and the actual observations. In the core region of the galactic halos, the density profile has been found to follow the power law scaling with the galactocentric radius $r$, that is, $\rho(r)\propto r^{-\alpha}$ \cite{deBlok:2009sp,Mikheeva:2007kp}. The computer simulations based on the CDM model predict that $\alpha \gtrsim 0.8$ \cite{Navarro:1995iw,Navarro:1996gj,Moore:1999gc,Klypin:2000hk,Taylor:2001bq,Colin:2003jd,Diemand:2005vz,Navarro:2008kc,Stadel:2008pn,Swaters:2003ApJ583732S}, while the observations suggest that $\alpha\sim 0.2$ instead \cite{deBlok:2009sp,Blok:2001fe,spekkens2005long}. (Note that some authors, based on CDM, found $\alpha\sim0.7$ instead \cite{Graham:2006ae}.)
In particular, the NFW profile \cite{Navarro:1995iw,Navarro:1996gj} with $\alpha=1$ (cusp) at the center of the galactic halo has been found to agree with the result of CDM simulations \cite{Navarro:1995iw,Navarro:1996gj,Taylor:2001bq,Colin:2003jd}. The pseudo-isothermal profile with $\alpha=0$ (core) at the center, on the other hand, fits the observation data with a smaller $\chi^2$ than NFW does. Evidently, a core is preferred over a cusp. This implies that if the dark matter is truly dust-like and if the CDM simulations are correct, then there would be an under-supply of gravity near the galactic center.

One can entertain multiple explanations to this cusp-core problem. For example, at small scales the galactic dynamics may become so complex that the straight-forward CDM ceases to be valid. That means it is likely that the current state of CDM simulations may not have included additional physics, such as baryon physics \cite{0004-637X-560-2-636,:2006tm,0004-637X-691-2-1300,mashchenko2006removal,sergey2008stellar,ricotti2004origin,Chen:2008xpa}, necessary for addressing this problem. On the other hand, the cusp-core problem may indeed be a signal that reveals the deficiency of CDM model, despite its great success at large scales. In this paper we choose to take the latter approach.

Within the framework of the CDM model, the rotation curve is directly related to the density profile through the relation \cite{Blok:2001fe},
\begin{align}
4 \pi G\rho(r)= 2\frac{v}{r}\frac{\partial v}{\partial r}+(\frac{v}{r})^2.
   \label{eq:one}
\end{align}
One can see that under the CDM model, a power-law density profile implies a power-law rotation curve, i.e., $v\propto r^\beta$, where $\alpha=2\beta-2$. Therefore the mismatch of the density profiles in the CDM cusp-core problem is equivalent to the mismatch of the rotation curves, which is what one obtains directly from observations. According to Eq.(1), there is a one-to-one correspondence between the rotation curve and the density profile under CDM model. In principle, therefore, there exists a CDM density profile solution for every given rotation curve. However, computer simulations based on CDM model can only produce the cusp density profile, which does not conform with the observed rotation curve near the center.

Several attempts have been made to build alternative DM models with non-zero pressure. Models considering dark matter as an ideal fluid with isotropic pressure is known to be inconsistent with observations. Models with anisotropic pressure, therefore DM behaves as a nonideal fluid, have been investigated and shown to be self-consistent \cite{Bharadwaj:2003iw,Su:2009fd,Matos:2000ki}. It has been suggested that scalar fields can be a good candidate for this type of dark matter. Several scalar field dark matter (SDM)models have been proposed \cite{Matos:2000ki,Arbey:2006it,He:2007tt,Lee:1995af,1475-7516-2008-05-005,1475-7516-2008-10-023}, although their considerations are not same as ours.

The behavior of such scalar field dark matter, however, is greatly dependent on the form of its self-interaction potential. Among the various proposals, the exponential potential is often considered \cite{Copeland:1997et}. Matos, Guzm\'{a}n and Nu\~{n}ez (MGN), for example, showed that a consistent solution of Einstein equation giving rise to a flat rotation curve exists, if the self-interaction potential is of the form of an exponential function. Such solution reproduces the rotation curve in the outer region of the galactic halo. \cite{Matos:2000ki,Nandi:2009hw} Matos et al. also showed that their model could well explain other observations such as that of type Ia supernovae \cite{Matos:1999et}. The exponential potential scalar field may find its theoretical connections with, for example, the radion or dilaton field in higher-dimensional gravity theories \cite{Green:1987sp,Carroll:2004st}. It can also be found in higher-order gravity theories \cite{Barrow:1988xh,Wands:1993uu}. In addition, the exponential-potential scalar field can be induced by non-perturbative effects such as gaugino condensation \cite{Carlos:1992da}. It is therefore quite natural to wonder, would such nonconventional SDM be able to solve the existing cusp-core problem associated with the conventional CDM model?

In this paper, we consider a novel scalar field with a potential $V(\Phi)=-a\exp(-b|\Phi|)$ as dark matter. This potential is essentially the same as that in MGN: $V(\Phi)=-a\exp(-b\Phi)$ except that we take the absolute value of the field in the exponent, so as to ensure the field is bounded from below in the system. We demonstrate that such a novel scalar field DM can readily reproduce the profile of the observed rotation curve from the galactic core to its outer region where the rotation curve is flat. Furthermore, the corresponding $\chi^2$ of our model is in general smaller than that of the NFW profile but comparable to that of the pseudo-isothermal  (ISO) profile. It is noteworthy that, by invoking only two parameters, which is the case in NFW and ISO, we are able to resolve the problem. That is, the exponential-potential SDM model is able to solve the existing cusp-core problem associated with the conventional CDM model in a simple construction. In the CDM model the rotation curve cannot be uniquely determined owing to the degeneracy of the Einstein equation under zero pressure \cite{deBlok:2009sp}. The determination of the rotation curve should therefore rely on simulations. In contrast, the rotation curve and the profile of energy-momentum tensor can be uniquely determined in our SDM model once the potential of the scalar field is given. Therefore, further simulation under our model is unnecessary once the unique solution that fits the observation is found.

The organization of this paper is as follows. In Sec. II, we introduce the basic setting of the galactic model proposed by Matos, Guzm\'{a}n and Nu\~{n}ez. But instead of a constant, we invoke a hyperbolic tangent function for the rotation curve, which rather faithfully represents the observed generic rotation curve as a function of the galactic radius. This represents the first step of our attempt to solve the cusp-core problem in one stroke. We then show that the SDM with an exponential self-interaction potential can indeed induce the desired rotation curve. In the Sec. III, we numerically compute the rotation curve induced by an exponential-potential scalar field dark matter with suitable boundary conditions. The resulting rotation curves are shown to fit the observational data very well. We also include the baryonic energy density in the Einstein equation and show that such setting is self-consistent. In the last section we discuss the implications of our results.

\section{The self-interaction potential}
Assuming spherical symmetry of galactic mass distribution, which includes both DM and baryons, we write down the general form of the metric:
\begin{align}
ds^2= -e^{2\sigma(r)}dt^2+e^{2\lambda(r)}dr^2+r^2d\theta^2+r^2sin^2\theta d\phi^2.
\end{align}
We note that the actual mass distribution is not exactly spherically symmetric since the stellar distribution is often non-spherical. Nevertheless, spherical symmetry has been shown to be a good approximation in the calculation of the rotation curve \cite{PhysRevD.65.023511}.
To investigate the rotation curve governed by a scalar field DM, we write the Einstein equations as
\begin{align}
R_{\mu\nu}=8\pi G(T_{\mu\nu}^{tot}-\frac{1}{2}T^{tot}g_{\mu\nu}),
\end{align}
where the total energy-momentum tress tensor is contributed by the SDM and the baryon matter:
\begin{align}
T_{\mu\nu}^{tot}=T_{\mu\nu}^{SDM}+T_{\mu\nu}^{b}.
\end{align}
With the aid of the Klein-Gordon equation that governs the SDM,
 \begin{align}
\Phi^{;\mu}_{;\mu}-\frac{dV}{d\Phi}=0,
\end{align}
the energy-momentum tensor $T_{\mu\nu}^{SDM}$ can be expressed as
\begin{align}
T_{\mu\nu}^{SDM}=\Phi_{,\mu}\Phi_{,\nu}-\frac{1}{2}g_{\mu\nu}\Phi^{,\sigma}\Phi_{,\sigma}-g_{\mu\nu}V(\Phi).
\end{align} 
On the other hand, in our treatment of the baryonic energy-momentum tensor $T_{\mu\nu}^b$, which is predominantly contributed by the stellar mass, is an input that is determined through the observation of the galactic luminosity.
The Einstein equation then reads
\begin{align}
R_{\mu\nu}=8\pi G[\Phi_{,\mu}\Phi_{,\nu}+g_{\mu\nu}V(\Phi)+\text{baryonic terms}].
\end{align} 
This is the only equation that governs our model.

Under spherical symmetry, the $g_{tt}$ component of the metric is directly related to the rotation velocity of the galactic halo through the equation \cite{Bharadwaj:2003iw}
\begin{align}
r\sigma(r)'=\Big[\frac{v(r)}{c}\Big]^2.
\end{align}
From Eqs.(7) and (8), we see that once the rotation curve is determined the number of independent equations is the same as the number of undetermined variables: $V(\Phi)$, $\Phi(r)$, $\sigma(r)$ and $\lambda(r)$.

In their model, Matos, Guzm\'{a}n and Nu\~{n}ez assumed the flatness of the rotation curve, with which they obtained the form of $\sigma(r)$ through Eq.(8). Inserting it into the Einstein equation, they found $V(\Phi)$ to be an exponential function \cite{Matos:2000ki}.

To investigate the property of SDM near the galactic core, where the rotation curve is no longer flat, we replace the constant by a hyperbolic tangent function to model the rotation curve for the entire range of radius, i.e.,
\begin{align}
v(r)=v_0\tanh(\beta r^\alpha),
\end{align}
where the radius $r$ is in units of kpc. This rotation velocity follows a power-law function at small $r$ but saturates to a constant, $v_0$, at large $r$. Such function fits the rotation curve reasonably well as shown in Fig. 1. When inserting the corresponding form of $\sigma(r)$ in the Einstein equation, we keep only the leading order terms of $\sigma(r)$, $\lambda(r)$ and $v^2/c^2$ in order to streamline the calculation. We note that such simplification is reasonable since the rotation velocity is small compared with the speed of light and that the gravity inside the halo is weak \cite{Bharadwaj:2003iw}. In particular, it can be verified that the Matos-Guzm\'{a}n-Nu\~{n}ez solution remains unchanged under such approximation.

\begin{figure}[H]
\center
  \includegraphics[width=8cm]{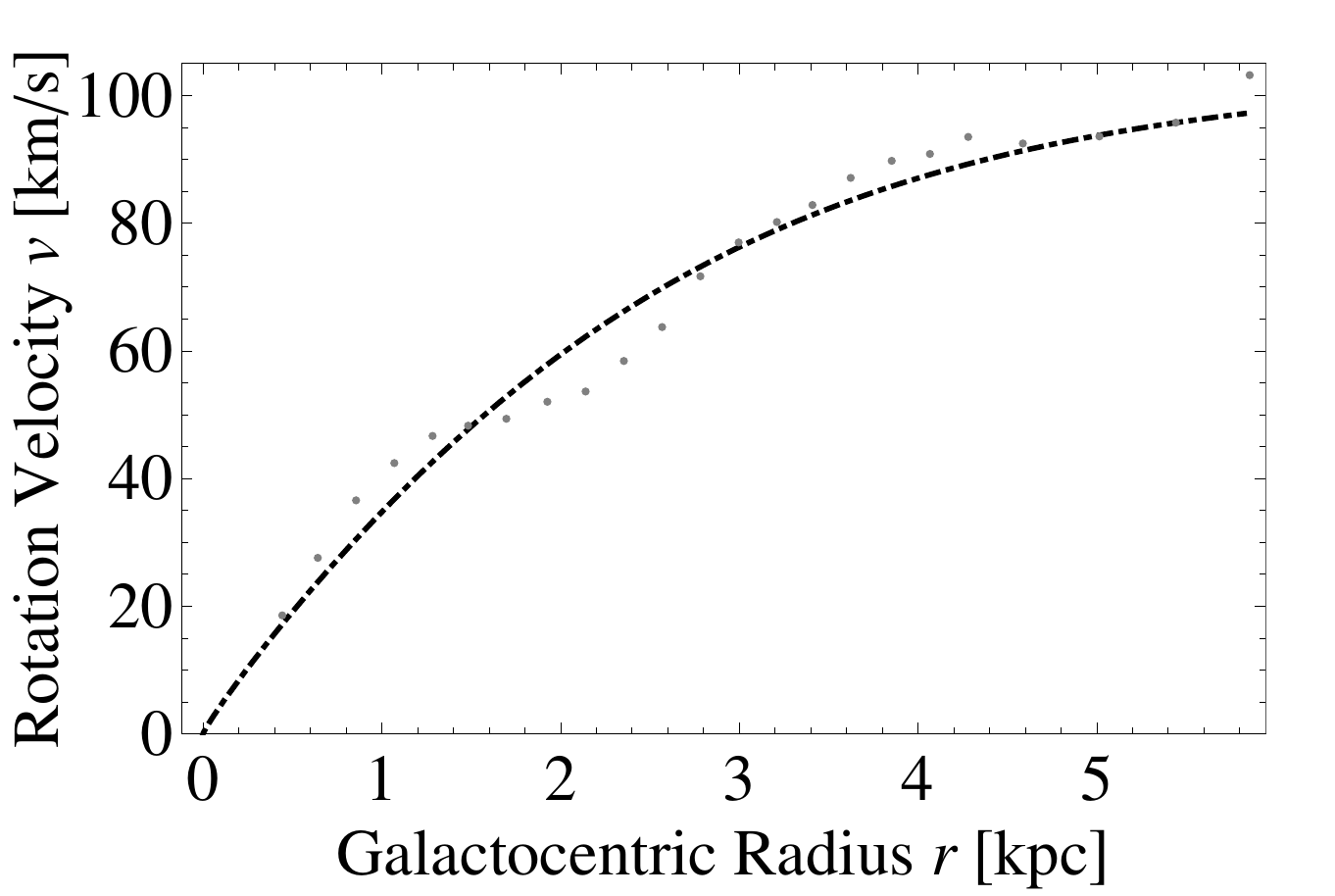}\\
  \caption{Using the hyperbolic tangent function to fit the dark matter contribution of the rotation curve of NGC 6689.}
\end{figure}

Neglecting the baryonic contribution for the time being, the Einstein equation reads:
\begin{align}
\frac{\lambda'}{r}-\frac{2\alpha v_0^2d}{c^2}r^{\alpha-2}{\rm sech}^2 (&dr^\alpha)\tanh(dr^\alpha)\notag\\
&=\frac{1}{\kappa}[(1+2\lambda)V(\Phi)+\frac{1}{2}\Phi'(r)^2]\notag\\
\frac{2\lambda}{r}+\frac{2\lambda}{r^2}&=\frac{1}{\kappa}[(1+2\lambda)V(\Phi)+\frac{1}{2}\Phi'(r)^2]\notag\\
\frac{-2\lambda}{r^2}+\frac{2v_0^2}{c^2r^2}\tanh^2(d r^\alpha)&=\frac{1}{\kappa}[-(1+2\lambda)V(\Phi)+\frac{1}{2}\Phi'(r)^2],
\end{align}
where $\kappa=1/8\pi G$. These equations are still too tedious to render simple analytic solution. However if we solve them numerically, we see, as shown in Fig. 2, that the actual solution for the potential comes very close to an exponential function. This gives us confidence that the exponential potential may indeed generate the desired rotation curve under suitable boundary conditions.

\begin{figure}[H]
\center
  \includegraphics[width=8cm]{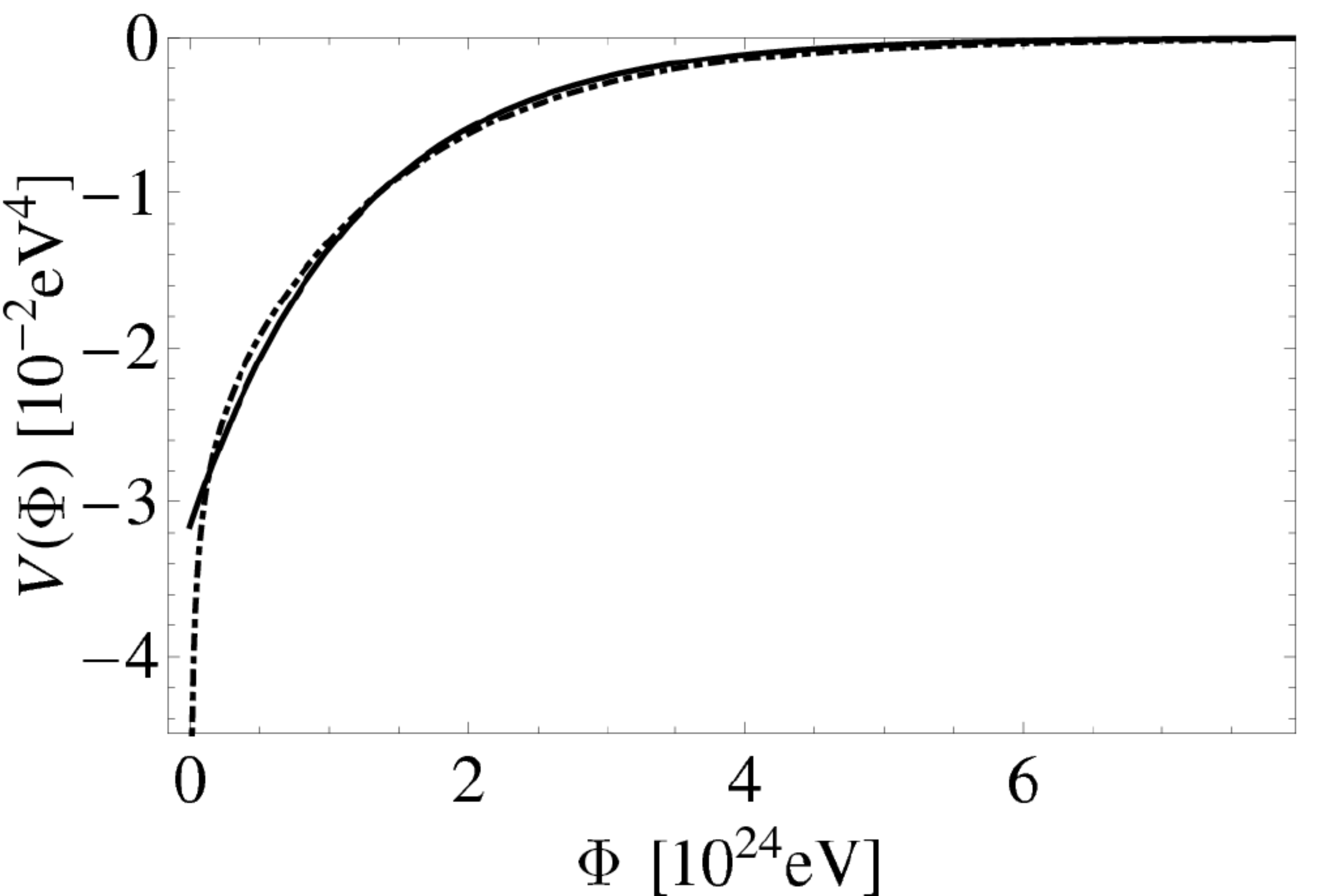}\\
  \caption{Numerical solution for the potential $V(\Phi)$ in Eq.(10). The values of $v_0$, $d$ and $\alpha$ used in this calculation are taken from the fitting of the rotation curve data of NGC 6689. The dot-dashed curve is the numerical solution of $V(\Phi)$ while the solid curve is the exponential fit.}
\end{figure}


\section{consequences of the exponential potential}
\subsection{Rotation Curve}
Guided by the numerical solution, we now turn around and assume that the SDM potential is indeed exponential: $V=-a\exp(-b\Phi)$, where $a$ and $b$ are positive. We substitute it into the Einstein equations and solve for the rotation curve. Ignoring the higher order terms as we did in Eq.(10), we have
\begin{align}
\frac{2\lambda}{r^2}+\frac{2\lambda'}{r}&=\frac{1}{\kappa}[V(\Phi)(1-2\lambda)+\rho_b(r)+\frac{1}{2}\Phi'(r)^2],\notag\\
-\frac{\sigma'}{r}+\frac{\lambda'}{r}-\sigma''&=\frac{1}{\kappa}[V(\Phi)(1-2\lambda)+\frac{1}{2}\Phi'(r)^2],\notag\\
-\frac{2\lambda}{r^2}+\frac{2\sigma'}{r}&=\frac{1}{\kappa}[-V(\Phi)(1-2\lambda)+\frac{1}{2}\Phi'(r)^2],
\end{align}
where $\rho_b$ is the baryon density.

\begin{figure}[H]
  \centering
  \includegraphics[width=7cm]{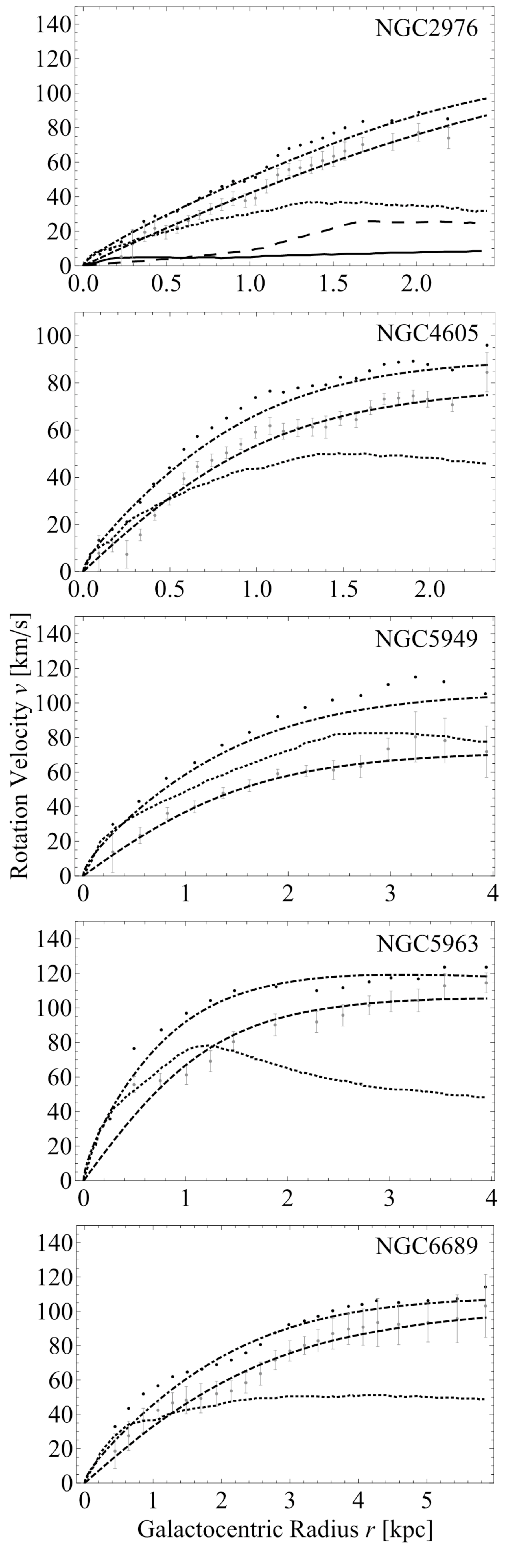}\\
  \caption{Comparisons of the numerical solutions with the observational data. The black dots are the observed rotation curves. The gray dots are the contribution of the dark matter, and the dotted curve is the contribution of the stellar disk. The dashed curve is our result that includes only the dark matter, while the dot-dashed curve is our result with the baryonic contribution taken into account. For NGC 2976 the long dashed curve represents the contribution of HI and the solid curve represents the contribution of $H_2$ }
  \end{figure}

The task of obtaining an explicit solution to these equations is still formidable. We therefore resort again to the numerical solution. We observe that in every term of these equations the lowest order of the $\sigma$ derivatives is one. Thus we only need to specify the boundary conditions of $\lambda$, $\sigma'$ and $\Phi$. We find that if the boundary conditions of all these three quantities are zero when $r$ is extremely small, then the resulting rotation curve has the same shape as that obtained from observations. Furthermore, the resulting $\Phi(r)$ is always positive definite in our model, the same as in MGN. This indicates that our replacement of $-a\exp(-b\Phi)$ by $-a\exp(-b|\Phi|)$ does not alter any basic notions of MGN except to insure that the quantum states are bounded from below.

Based on this solution, we analyze the five galaxies from Simon et al. \cite{Simon:2003xu,Simon:2004sr} and two of the 19 galaxies from Blok et al. \cite{deBlok:2008wp} whose $\chi^2_{NFW}$ surpasses $\chi^2_{ISO}$ the most, meaning that in these samples cores are greatly preferred over cusps. Among all of the analyzed galaxies, NGC 5963 is the only one with $\chi^2_{SDM}$ larger than $\chi^2_{NFW}$.

Simon et al. \cite{Simon:2003xu,Simon:2004sr} have considered both the maximum and the minimum disks. We note that in their analysis of the maximum disk, they subtract the contribution of the stellar disk from the observed rotation curve before they fit it to the NFW and the pseudo-isothermal profiles. In their analysis of the minimum disk, however, they fit the observed rotation curves directly without subtracting the stellar disk contribution.
Therefore we choose to compare our model with the other two mass models under the assumption of maximum disk only, which, as commented by Simon et al., is more physical. Before doing that we first use Eq.(8) and Eq.(11) without including the baryonic density profile to calculate the SDM contribution alone. The comparison of $\chi^2_{SDM}$ and the $\chi^2$ of the other two profiles are shown in Table.1, from which we see that except NGC 5963, all $\chi^2_{SDM}$ are smaller than $\chi^2_{NFW}$ and comparable to $\chi^2_{ISO}$. In the case of NGC 4605, $\chi^2_{SDM}$ is even smaller than $\chi^2_{ISO}$. Comparisons of our numerical solutions to the rotation curve with the observational data are shown in Fig.3.

In our calculation of the total rotation curve, we include the stellar disk contribution by invoking the fitting function, $v_b^2(r)=1.97 v_b^2(r_{opt})(r/r_{opt})^{1.22}/\left[(r/r_{opt})^2+0.78^2\right]^{1.43}$, which is commonly used \cite{Arbey:2006it,persic9506004mnras}. However the gas contribution differs from galaxy to galaxy, so in our actual calculation we used a four-to-six term polynomial to spline the discrete data points. We see from Fig. 3 that both results, the inclusion and the exclusion of the stellar disk contributions, fit the data reasonably well. Note that in order not to make the graphs too busy, only the error bars of the dark matter contribution are included. After all, that is what we need for the calculation of the $\chi^2_{SDM}$ under the maximum disk assumption.

In the paper of Blok et al. \cite{deBlok:2008wp}, the data was analyzed through a different method. They considered two cases where the $\gamma_\star^{3.6}$, that is, the light-to-mass ratio at $3.6\mu m$, is either fixed through the mass model or kept as a free parameter. They then considered two probable descriptions of the dark matter distribution, the NFW profile and the pseudo-isothermal profile. The total rotation curve, which is the square root of the quadratic sum of the rotation curves contributed by all the components, is then fitted to the observed rotation curve from which the corresponding $\chi^2$ is calculated.

\begin{figure}[H]
  \centering
  \includegraphics[width=7cm]{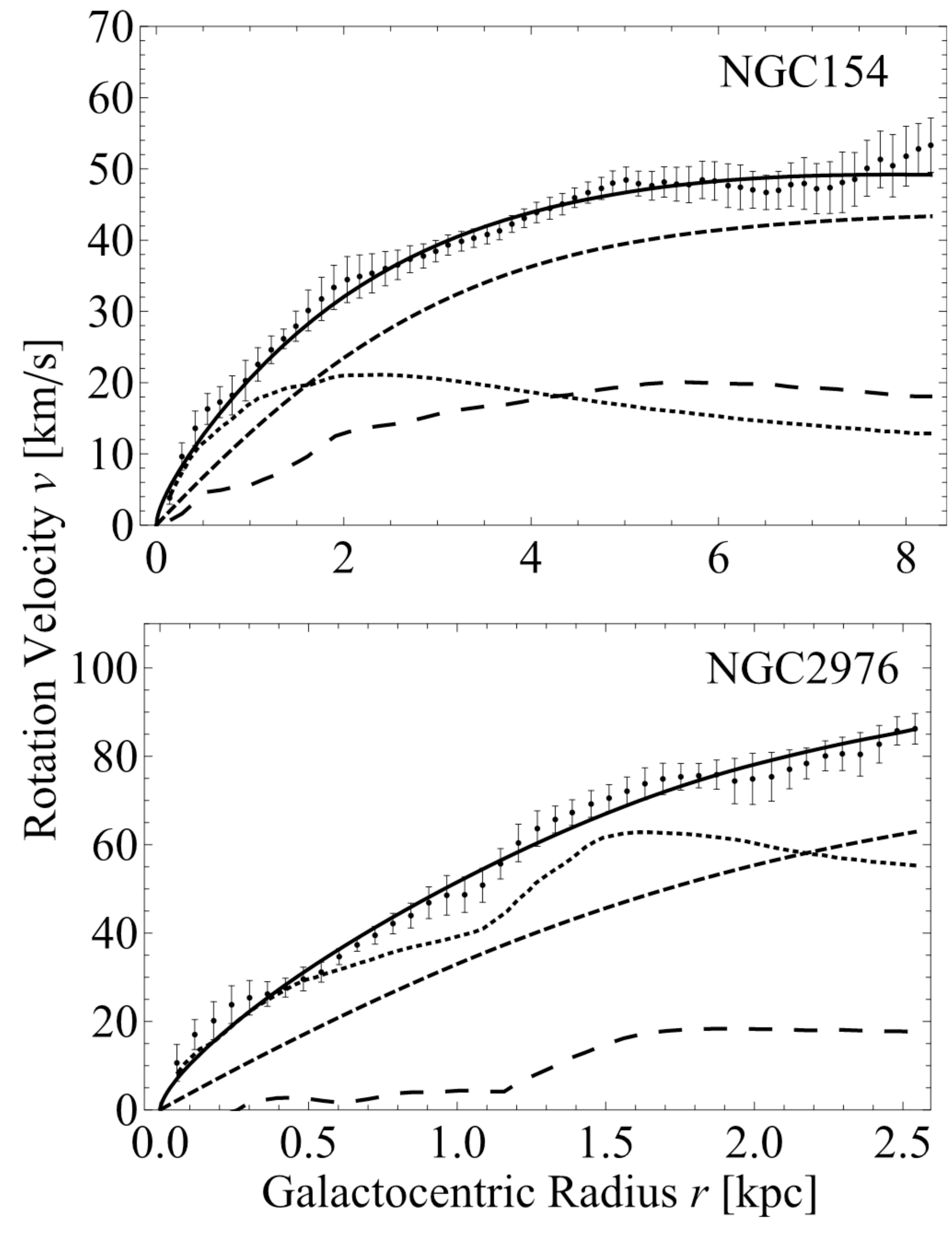}\\
  \caption{Comparisons of the numerical solutions with the rotation curves based on the observational data from Blok et al. \cite{deBlok:2008wp}. The black dots are the observed rotation curves, the dotted curve is the contribution of stellar disk, and the long dashed curve is the contribution of gas. The black curve is our result that takes the baryonic contribution into account. The dashed curve is our result that considers only the dark matter.}
  \end{figure}

For the two galaxies from Blok et al. \cite{deBlok:2008wp}, we assume that $\gamma_\star^{3.6}$ is the value that minimizes the $\chi^2_{ISO}$ and use Eq.(11) and Eq.(8) to calculate the total rotation curve and the corresponding $\chi^2_{SDM}$. We find that although we do not treat $\gamma_\star^{3.6}$ as a free parameter, we readily obtain a $\chi^2_{SDM}$ that is much smaller than $\chi^2_{NFW}$.   Comparisons of our numerical solutions to the rotation curves with the observational data from Blok et al. \cite{deBlok:2008wp} are shown in Fig.4. Once again, we find that our solutions fit the observations quite well.

\begin{table}
\caption{\label{tab:table1}$\chi^2$ of the data from Simon et al. under maximum disk.}
   \begin{ruledtabular}
  \begin{tabular}{cccc}
$\chi^2$\footnote{$\chi^2$ per degree of freedom.} & NFW & Pseudo-Isothermal & SDM\\
\hline
NGC 2976 & $>5.8$  & $0.43$ & $\lesssim 0.44$ \\
NGC 4605 & $<8.03$ & $1.53$ & $\lesssim 1.23$ \\
NGC 5949 & $<1.24$ & $0.24$ & $\lesssim 0.30$ \\
NGC 5963 &  $0.43$ & $1.46$ & $\lesssim 2.01$ \\
NGC 6689 & $<1.31$ & $0.46$ & $\lesssim 0.50$ \\
  \end{tabular}
  \end{ruledtabular}
\end{table}

\begin{table}
\caption{\label{tab:table1}$\chi^2$ of the data from Blok et al. }
   \begin{ruledtabular}
  \begin{tabular}{cccc}
$\chi^2$\footnote{$\chi^2$ per degree of freedom.} & NFW \footnote{Free $\gamma_\star^{3.6}$}  & Pseudo-Isothermal\footnote{Free $\gamma_\star^{3.6}$}  & SDM\footnote{$\gamma_\star^{3.6}$ minimizing $\chi^2_{ISO}$} \\
\hline
NGC 2976 & $1.65$  & $0.51$ & $\lesssim 0.64$ \\
DDO 154  & $0.81$  & $0.28$ & $\lesssim 0.39$ \\
  \end{tabular}
  \end{ruledtabular}
\end{table}

\subsection{Energy-Momentum Tensor and Energy Conditions}
Having the desired rotation curve successfully deduced from the SDM exponential potential, we  now inspect the property of the corresponding energy and pressure density profiles of the SDM halo. The density and the pressure profiles of the dark matter can be expressed in terms of the potential and the field via Eq.(6):
\begin{align}
\rho_{dark}&=\frac{1}{2}e^{2\lambda(r)}\Phi'(r)^2+V(r),\notag\\
p_r&=\frac{1}{2}e^{2\lambda(r)}\Phi'(r)^2-V(r),\notag\\
p_t&=-\frac{1}{2}e^{2\lambda(r)}\Phi'(r)^2-V(r).
\end{align}
It is evident from these equations that the SDM model satisfies the non-ideal fluid equation of state, $\rho_{dark}(r)=-p_t(r)$.

We again take NGC 4605 as an example. Fig.5 shows the energy density profiles, where $\rho_{dark}$ corresponds to the SDM halo and $\rho_{tot}$ includes both DM and the baryonic matter. The radial and tangential components of the pressure are shown in Fig. 6. From these figures we see that the density and the pressures are of the same order of magnitude. This implies that the two terms, including the potential, on the RHS of Eq.(12) are also having the comparable magnitude.
\begin{figure}[h]
  \includegraphics[width=8cm]{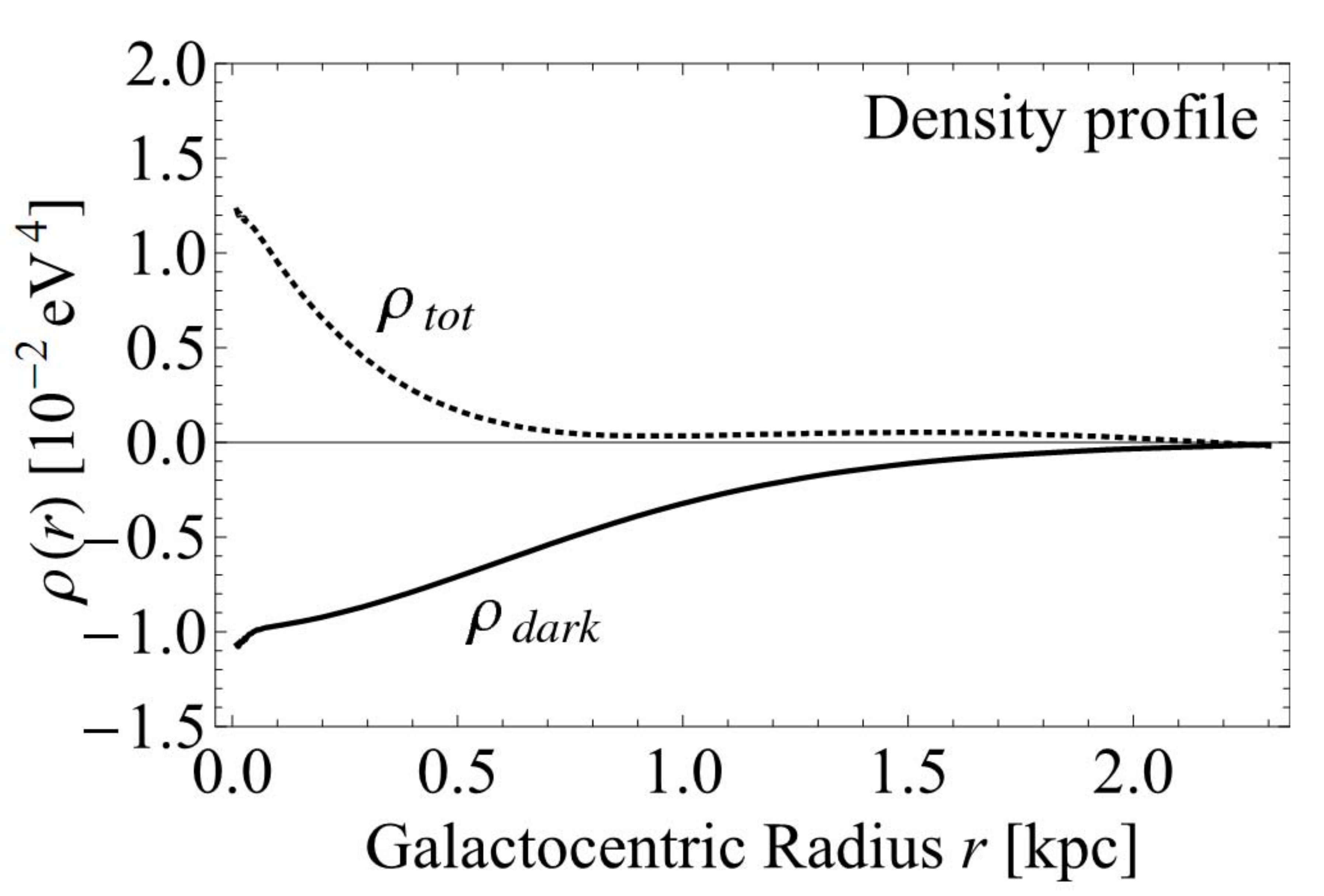}\\
  \caption{The solid is the density profile for the dark matter. The dotted curve is the density profile considering both the baryon and dark matter.}
\end{figure}

\begin{figure}[h]
  \includegraphics[width=8cm]{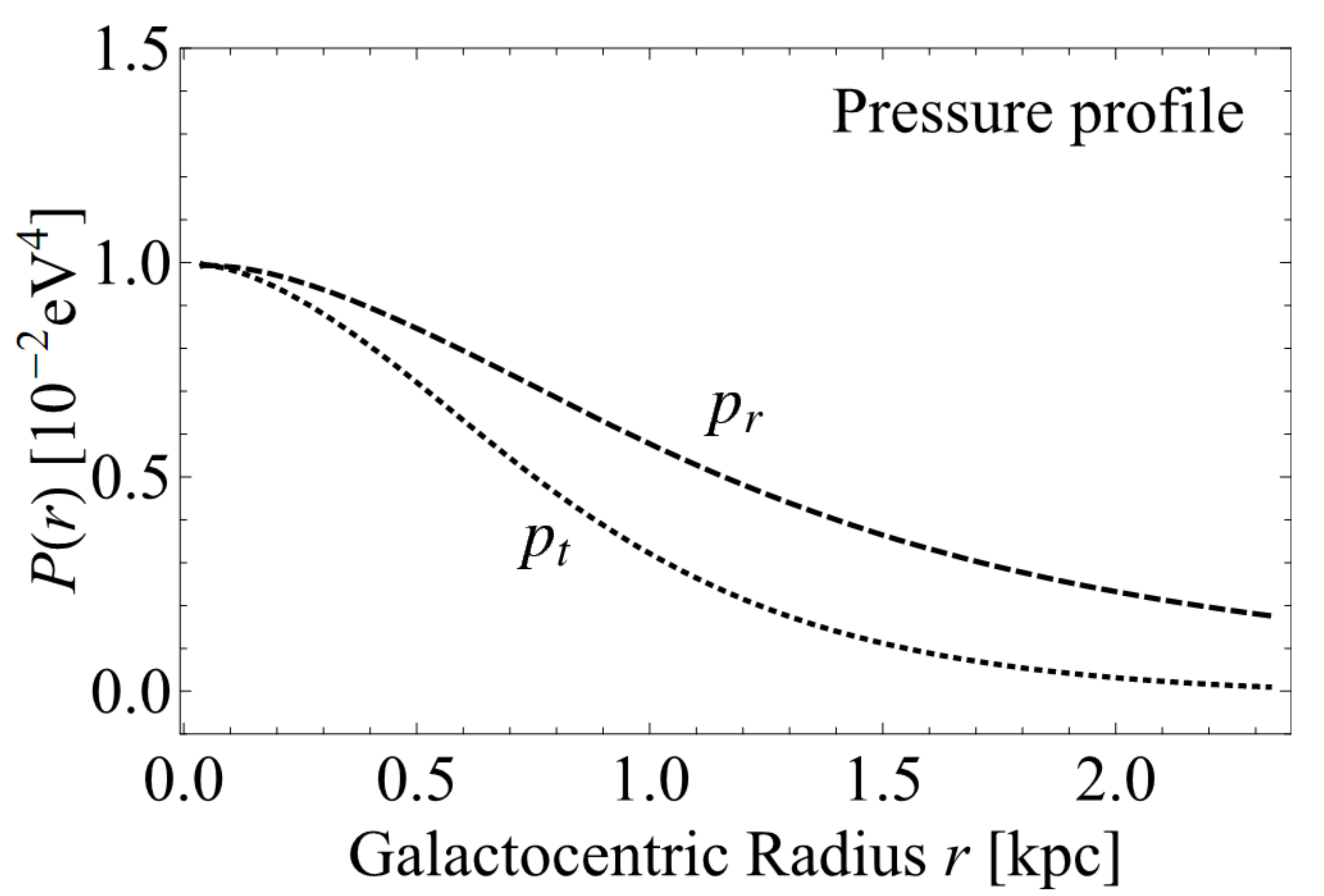}\\
  \caption{The dotted curve is the profile of tangential pressure. The dashed curve is the profile corresponding to the radial part of the pressure.}
\end{figure}

We see that although the energy density profile of the SDM is negative, the total energy density that includes the contribution from the baryons is always positive definite and therefore the system satisfies the weak energy condition \cite{Carroll:2004st}. We note that the dark matter sector itself, though violating the weak energy condition, still satisfies the null and the strong energy conditions \cite{Carroll:2004st}. We therefore consider our solution reasonable. In addition to NGC 4605, we analyzed five additional objects: DDO154, NGC 2976, NGC 5949,NGC 5963, and NGC 6689. Again the null energy condition and the strong energy condition are satisfied in the region of interest, although the weak energy condition is not always satisfied.

A negative density profile may seem unphysical at first sight. However, this is not true. In the conventional CDM model, owing to the fact that it provides no pressure, the energy density profile must be positive so as to render a positive, or attractive, gravity, which is the only required property for DM based on observations. For galactic models that are semi-Newtonian, this requirement remains unchanged. However, in our model the magnitude of the pressure is comparable to that of the energy density. This changes the situation drastically. That is, in our case the strong energy condition ensures the attraction of the gravity without relying on the positivity of the energy density, thanks to the nontrivial and substantial contribution of the pressure.

To be more specific, it can be seen in the work of Faber and Visser \cite{Faber:2005xc} that the effective mass distributions associated with the rotation curve and the gravitational lensing are
\begin{align}
m_{RC}(r)=\int dr 4\pi r^2\left[ \rho(r)+p_r(r)+2p_t(r) \right]
\end{align}
and
\begin{align}
m_{lens}(r)=\int dr 4\pi r^2\left[ \rho(r)+ \frac{1}{2} (p_r(r)+2p_t(r))\right]\,
\end{align}
respectively. In our model both $\left[ \rho(r)+p_r(r)+2p_t(r) \right]$ and $\left[ \rho(r)+ \frac{1}{2} (p_r(r)+2p_t(r))\right]$ are positive definite for all $r$, which ensures that our effective mass distributions are positive and that the gravity is attractive.

We note that negative energy density profiles do exist in physics. For example the energy density of Casmir energy is often negative. There also exist dark matter models that imply a negative energy density profile \cite{Matos:2000ki,Arbey:2006it}. In the model of Arbey et al., a repulsive gravity in some region was even considered.

In the conventional CDM model the rotation curve is solely governed by the DM density profile through Eq.(1). In our model where the pressure is nonzero and anisotropic, the energy-momentum that drives the stellar rotation velocity $\rho(r)$ is replaced by $\rho_{dark}(r)+p_r(r)+2p_t(r)$ \cite{Faber:2005xc}. That is,
\begin{align}
4 \pi G [\rho_{dark}(r)+p_r(r)+2p_t(r)]= 2\frac{v}{r}\frac{\partial v}{\partial r}+(\frac{v}{r})^2.
   \label{eq:one}
\end{align}
However, we also know that in our model $\rho_{dark}(r)=-p_t(r)$. This means that the rotation curve is determined by the sum of the two pressure components, $p_r+p_t$ (but {\sl not} the total pressure $p_{tot}\equiv \sqrt{p_r^2+p_t^2}$, however). It is interesting to note that while in CDM the rotation curve is dictated entirely by the density, in contrast in SDM it is entirely governed by the pressure. As for the relative contributions between the two pressure components, we see from Fig.5 that both the radial and the tangential pressures exhibit similar radial dependence with roughly the same magnitude, except that the radial component is consistently a few times larger than that of the tangential one over essentially the entire range of the radius.

Evidently the derivative of our potential, $-a\exp(-b|\Phi|)$, is discontinuous at $\Phi=0$ and the model is therefore ill-defined. On the other hand, we found that the resulting rotation curve in our model is totally insensitive to the the profile of potential around $\Phi=0$. One may therefore regard our potential as a phenomenological model that is valid for all values of $\Phi$ except 0. We assume that the true potential is the combination of this exponential potential and a small correction term that rounds off its sharp edge at $\Phi=0$. Being minute and effective only at $\Phi\sim 0$, such addition to the potential would not affect any of our profile analysis.

\section{Conclusion}
 We have shown that the scalar field dark matter with an exponential self-interacting potential can successfully generate a unified and universal stellar rotation curve that covers from the galactic core to its outer region. Moreover, the $\chi^2_{SDM}$ based on our scalar dark matter model is found to be generally comparable with $\chi^2_{ISO}$ and smaller than $\chi^2_{NFW}$.  In some of the galaxies, we even found $\chi^2_{SDM}$ smaller than $\chi^2_{ISO}$. The only galaxy with $\chi^2_{SDM}>\chi^2_{NFW}$, i.e., NGC 5963, may be an exceptional case as commented by Blok et al. The brightness of its inner disk indicates that it may not be dark matter dominated all the way to the center \cite{deBlok:2009sp}. We recall that the pseudo-isothermal profile (ISO) agrees with the observation data whereas the NFW profile, which conforms with the CDM-based simulations, does not agree with the observation data near the galactic center. Being more consistent with the ISO profile and thus the observations, the scalar dark matter (SDM) model ameliorates the cusp-core problem associated with the CDM model.

In the conventional CDM model, there are in general three unknowns to be solved in the Einstein equations under the assumption of spherical symmetry. These are the DM density profile $\rho_{dark}(r)$ and the two metrics $g_{tt}(r)$ and $g_{rr}(r)$. However, since the CDM has no pressure, the Einstein equations are degenerate and that renders only two independent equations in the leading order expansion \cite{Bharadwaj:2003iw}. As a consequence, only the relationship between these three unknowns, instead of their absolute values, can be deduced. This is why it is not possible to derive the rotation curve from CDM model based on first principles. One usually relies on numerical simulations to discriminate possible solutions \cite{deBlok:2009sp}.
In our SDM model there are also three unknown quantities, $\Phi(r), g_{tt}(r)$ and $g_{rr}(r)$, to be determined after the form of $V(\Phi)$ is specified. However the difference is that in our case the pressure is nonzero and thus all three Einstein equations are non-degenerate. The three unknowns can then be determined explicitly. The profiles based on the SDM model are thus attainable from first principles without the need to resort to simulations. The core (instead of cusp) profile so deduced from SDM is therefore unambiguous. We find this aspect rather novel.

We have demonstrated the universality of the form of our SDM potential among the galaxies that we investigated, with $a\sim \mathcal{O}(10^{-2})$eV and $b\sim \mathcal{O}(10^{-24})({\rm eV})^{-1}$, although these values differ slightly from galaxy to galaxy. Comparing these values with that in MGN \cite{Matos:2000ki}, we find that the exponent of our potential comes very close to that in their model, where $b\sim 2\sqrt{\kappa c^2/v_0^2}$. Such correspondence is expected. MGN considered only the case where the rotation curve is a constant, which corresponds to our model at large $r$. If the rotation curve is constant for all $r$, the Einstein equation is analytically solvable, which has been carried out in MGN's work. Since our result should approach that of MGN at large $r$, the value of our $b$ is expected to be similar to theirs.

Although we mentioned in the introduction that the exponential potential is quite common in higher dimensional or higher order gravity theories, so far we are not able to deduce the coefficients in the exponential potential from first principles. Our model should therefore be viewed as a phenomenological or effective theory for dark matter halo. Even so, from the phenomenological point of view it is remarkable that by replying on no more parameters than that in NFW, our model can solve the cusp-core problem.

It is encouraging that our novel scalar field dark matter model with an exponential potential is able to resolve the cusp-core problem that stems from the cold dark matter model. Before this SDM model can be taken more seriously, however, it is essential that this non-conventional SDM can reproduce the large scale structure formation history of the universe, which is what the conventional SDM does its best. Another challenge is the need of an explicit SDM production scenario in the early universe based on a fundamental theory that would give rise to the desired amount of such SDM that would saturate the DM content of the universe.

\section*{Acknowledgement}
We thank S.H. Shao, Y. D. Huang and C. I. Chiang for interesting and inspiring discussions. This research is supported by Taiwan National Science Council under
Project No. NSC 97-2112-M-002-026-MY3 and by US Department of Energy under Contract No. DE-AC03- 76SF00515. We also thank the support of the National Center for Theoretical Sciences of Taiwan.

\bibliographystyle{REV}
\bibliography{halo}

\end{document}